\documentclass[11pt,twoside]{article}
\usepackage{graphicx,epsfig,natbib,epstopdf}
\usepackage{CS18}
\begin{document}
%
%
%
\title{A tale of two exoplanets: the inflated atmospheres of the Hot Jupiters HD 189733 b and CoRoT-2 b}

\author{K.\ Poppenhaeger$^{1, 3}$, S.J.\ Wolk$^{1}$, J.H.M.M.\ Schmitt$^{2}$}
\affil{$^1$Harvard-Smithsonian Center for Astrophysics, 60 Garden Street, Cambridge, MA 02138, USA}
\affil{$^2$Hamburger Sternwarte, Gojenbergsweg 112, 21029 Hamburg, Germany}
\affil{$^3$NASA Sagan Fellow}
\begin{abstract}

Planets in close orbits around their host stars are subject to strong irradiation. High-energy irradiation, originating from the stellar corona and chromosphere, is mainly responsible for the evaporation of exoplanetary atmospheres. We have conducted multiple X-ray observations of transiting exoplanets in short orbits to determine the extent and heating of their outer planetary atmospheres. In the case of HD 189733 b, we find a surprisingly deep transit profile in X-rays, indicating an atmosphere extending out to 1.75 optical planetary radii. The X-ray opacity of those high-altitude layers points towards large densities or high metallicity. We preliminarily report on observations of the Hot Jupiter CoRoT-2 b from our Large Program with XMM-Newton, which was conducted recently. In addition, we present results on how exoplanets may alter the evolution of stellar activity through tidal interaction.
\end{abstract}

\section{Exoplanetary transits in X-rays}

Close-in exoplanets are expected to harbor extended atmospheres and in some cases lose mass through atmospheric evaporation\index{evaporation}, driven by X-ray and extreme UV emission from the host star \citep[for exmaple]{2004A&A...418L...1L, 2009ApJ...693...23M}. Direct observational evidence for such extended atmospheres has been collected at UV wavelengths \citep{2003Natur.422..143V, 2010A&A...514A..72L, 2013A&A...551A..63B}. However, the constraints on hydrogen densities in the extended atmosphere are weak, and depending on the ionization degree, can vary over several orders of magnitude \citep{2010ApJ...709.1284B}. 

\begin{figure}[ht!]
\centering
\includegraphics[width=0.35\textwidth]{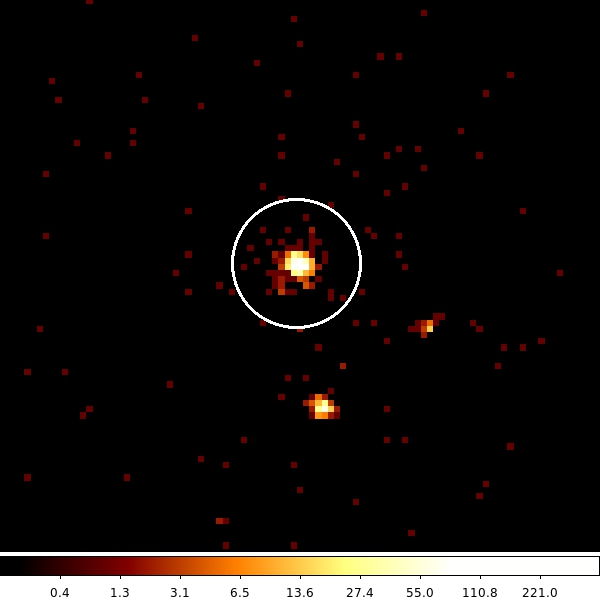}
\hspace{2cm}
\includegraphics[width=0.35\textwidth]{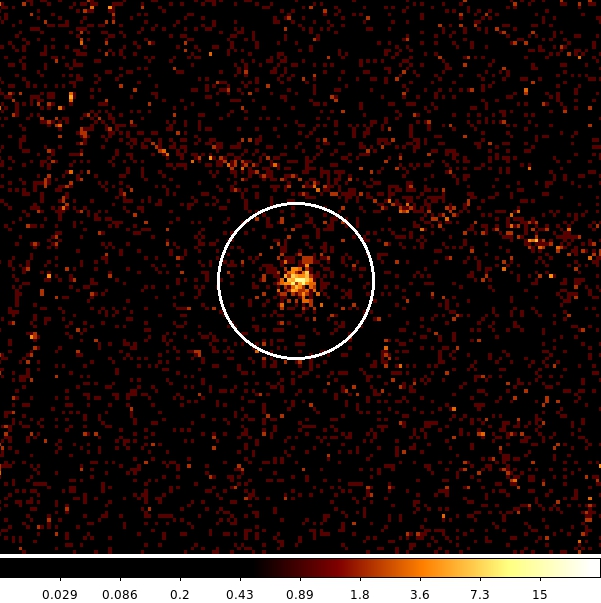}
\caption{X-ray image of HD 189733, observed with {\it Chandra} on the left, and CoRoT-2 observed with {\it XMM-Newton} on the right.}
\label{images}
\end{figure}

We have conducted observational campaigns of transiting exoplanets in the soft X-ray regime (0.2-2 keV), where opacity is not caused by hydrogen, but by heavier elements such as oxygen, nitrogen, and carbon. A growing number of transiting exoplanet hosts has been detected in X-rays now, for example GJ~1214 \citep{2014ApJ...790L..11L}, CoRoT-7 \citep{2012A&A...541A..26P}, and GJ~436 \citep{2010A&A...515A..98P}. For our present study we chose the very X-ray bright stars HD 189733 and CoRoT-2, both of which host a transiting hot Jupiter. Using the X-ray telescopes {\it XMM-Newton} and {\it Chandra}, we have repeatedly observed the transits of HD 189733\,b and CoRoT-2\,b. The X-ray images of the host stars are shown in Fig.~\ref{images}. In each single observation, the host star is clearly detected. 

\begin{figure}[ht!]
\centering
\includegraphics[width=0.48\textwidth]{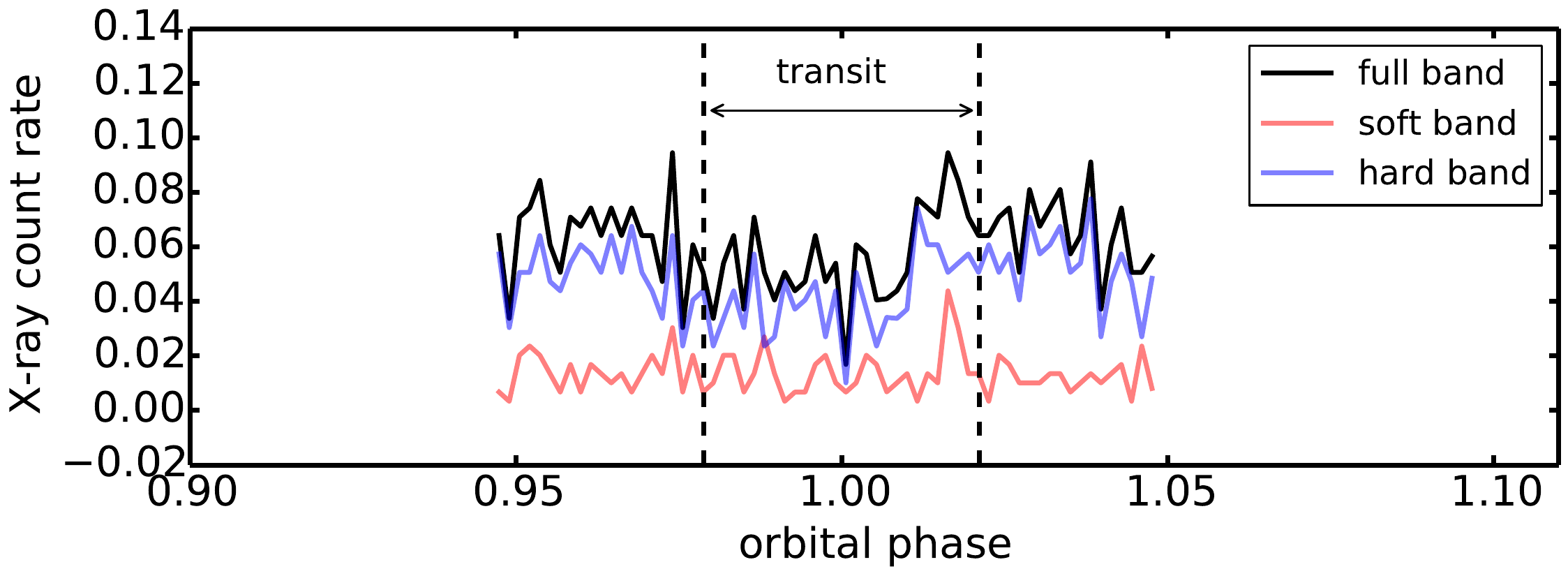}
\includegraphics[width=0.48\textwidth]{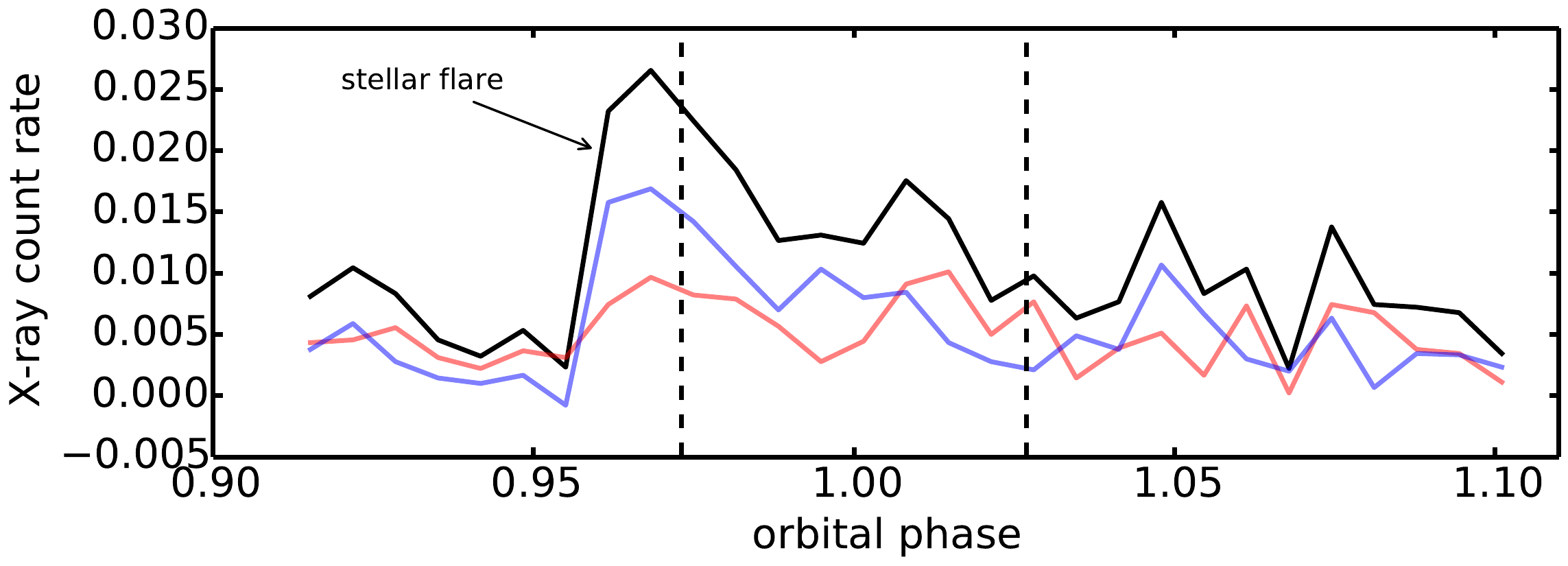}
\caption{Single X-ray light curves of HD 189733 (left) and CoRoT-2 (right) covering the planetary transits. CoRoT-2 displays a high level of intrinsic stellar variability.}
\label{single_lcs}
\end{figure}

We then extracted X-ray light curves from each individual observation. A typical example of a single light curve for each target is shown in Fig.~\ref{single_lcs}. In the case of HD 189733, we observed small-scale stellar variability, but large flares are rare. Several small flares have been observed at other orbital phases, namely after the exoplanetary eclipse \citep{2010ApJ...722.1216P, 2011ApJ...741L..18P, 2014ApJ...785..145P}. CoRoT-2 has been previously observed in X-rays \citep{2011A&A...532A...3S} with a high X-ray luminosity of $2\times 10^{29}$ erg/s. In our observations we confirm the high X-ray flux and additionally find several large flares; an example is shown in Fig.~\ref{single_lcs}, right. 

While the transit signal of the planet is not obvious in the individual light curves, we have detected the X-ray transit of HD 189733\,b in the stacked observations. We show the resulting co-added light curve in Fig.~\ref{transit_spin}, left. The X-ray transit\index{X-ray transit} is detected at a confidence of 99.8\%. The observed X-ray transit depth is 6-8\%, depending on the assumed model \citep{2013ApJ...773...62P}. We can theoretically expect a limb-brightened transit model with a "w"-like shape \citep{2010ApJ...722L..75S}, because the corona is optically thin and therefore limb-brightened (in contrast to the limb-darkened photosphere). However, the signal-to-noise is not high enough to reliably discriminate between limb-brightened and limb-darkened models. From the detected transit depth, we infer a minimum extent of the exoplanetary atmnosphere out to ca.\ 1.75 optical planetary radii, with particle number densities of the order of ca.\ $10^{11}$ cm$^{-3}$ when assuming an atmospheric metallicity of ten times solar. 

For CoRoT-2\,b, the individual light curves are much more dominated by intrinsic stellar variability, as the host star is more X-ray luminous by a factor of ten comapred to HD 189733. Detailed data analysis is still ongoing in order to identify suitable energy bands to minimize the influence of impulsive flares on the light curves.

\section{The X-ray luminosity of exoplanet host stars}

\begin{figure}[ht!]
\centering
\includegraphics[width=0.48\textwidth]{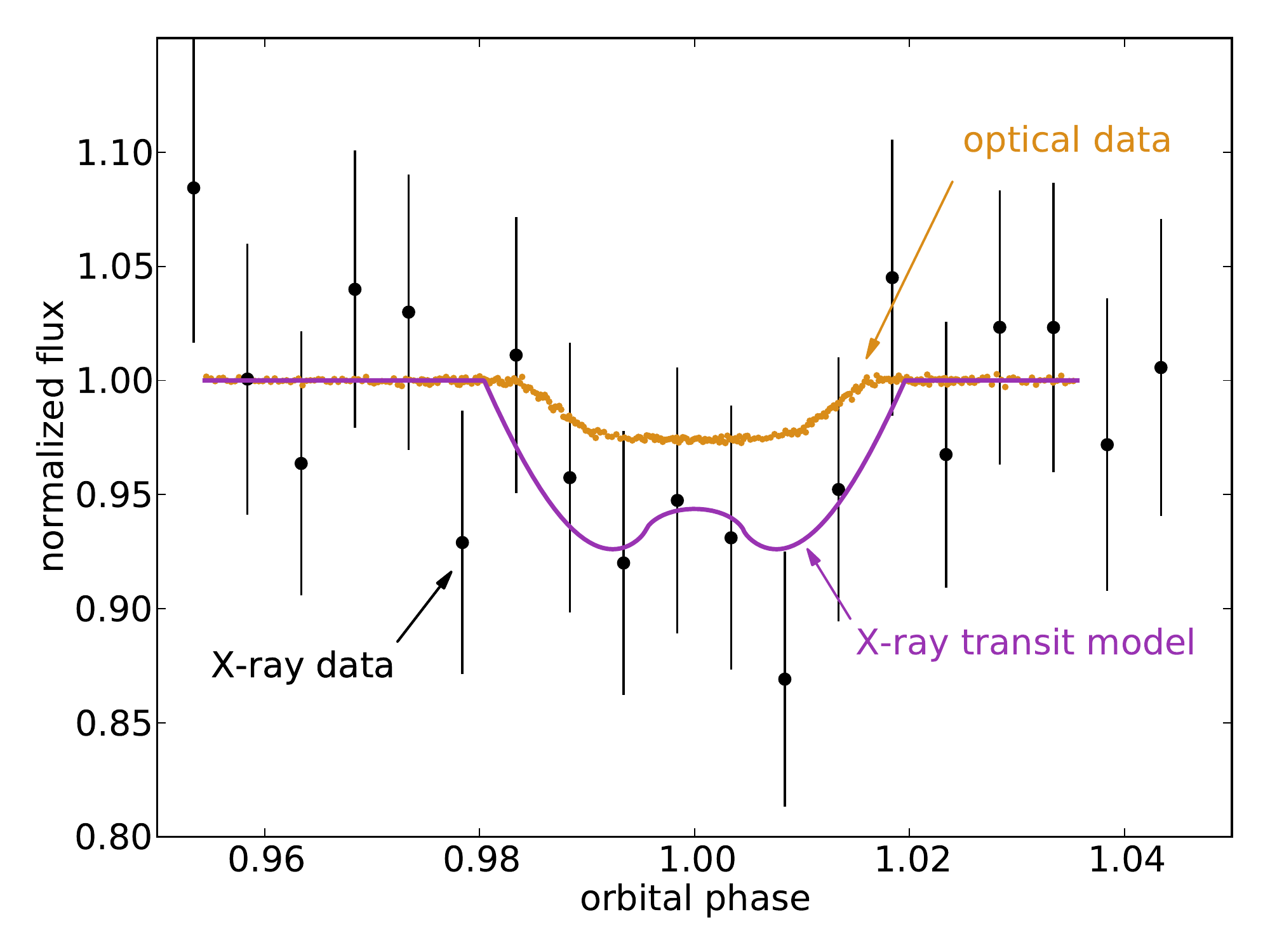}
\includegraphics[width=0.48\textwidth]{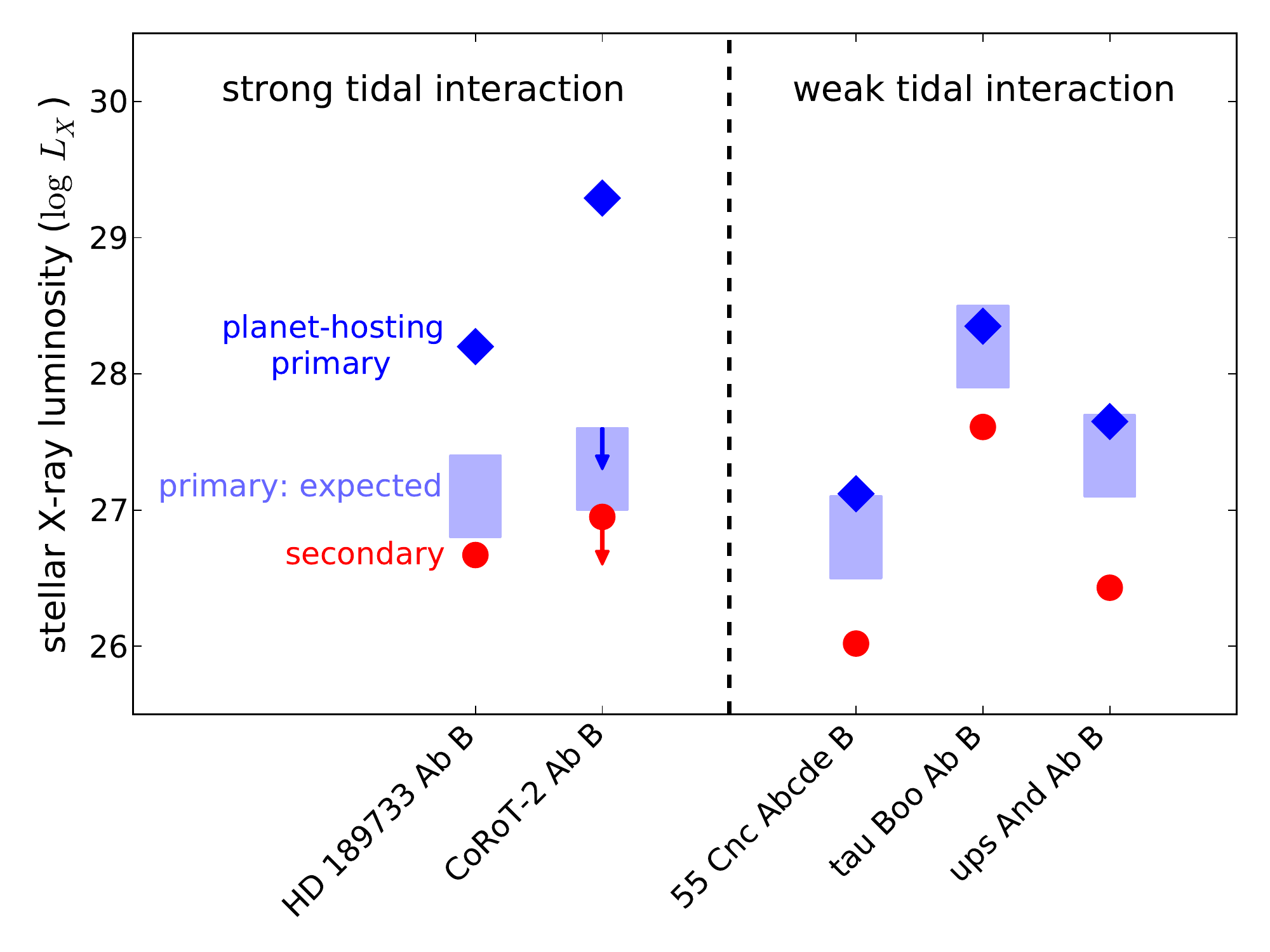}
\caption{{\rm Left:} Co-added X-ray light curves of HD 189733 reveal the transit signature in X-rays. The X-ray transit depth is ca. 3 times as large as the optical transit depth, indicating the presence of high-altitude atmosphere layers. {\rm Right:} X-ray luminosities of the stars in five wide binary systems, in which the primary is known to host an exoplanet. Assuming a common age of the two components in a system, we estimated the expected X-ray luminosity of the planet-hosting star (blue shaded area) based on the observed X-ray luminosity of the secondary star (red circles) and age-activity relationships. In the systems with expected strong tidal interaction between planet and star, the planet-hosting stars display much stronger X-ray emission (blue diamonds) than expected. }
\label{transit_spin}
\end{figure}

For X-ray studies of exoplanets it is helpful to have a relatively X-ray bright host star with low intrinsic variability. Magnetic activity\index{magnetic activity} is the main source of both variability and X-ray emission for cool stars, and it is an intriguing question if close-in exoplanets can actually have an effect on the stellar activity though tidal or magnetic star-planet interactions\index{star-planet interactions} \citep{2000ApJ...533L.151C}. Observational indications for such an (intermittent) magnetic effect on the chromospheres of exoplanet host stars have been reported early on for two systems \citep{2005ApJ...622.1075S, 2008ApJ...676..628S}. Indications for tidal effects have been found through observations of stellar rotation periods and velocities \citep{2009MNRAS.396.1789P, 2012MNRAS.422.3151H} as well as orbital obliquities of exoplanets \citep{2012ApJ...757...18A}. 

To exclude observational biases from distorting the sample of host stars and their activities, which can arise due to the fact that small exoplanets often go unnoticed around active stars \citep{2011ApJ...735...59P}, we have developed a new approach to test for a planet-induced high stellar activity level. We have selected wide (several 100 AU) stellar binaries\index{binaries} in which only one of the stellar components is known to host an exoplanet. We have then obtained X-ray observations of the two stars in each binary system. With a known spectral type, age-activity relationships \citep[for example]{2008ApJ...687.1264M, 2005ApJS..160..390P, 2011ASPC..451..285E} allow to test if the two stars display activity levels which are consistent with a common age.

We have used the X-ray luminosity of the secondary, which is the star without a known planet in our sample of five wide binaries, to estimate the age of the respective wide binary system. We then compare the observed X-ray luminosity of the planet-hosting star to the X-ray luminosity expected for that age. We find that for the two systems in which strong tidal interaction between exoplanet and host star is expected, the planet-hosting stars are over-luminous in X-rays by 1-2 orders of magnitude (see Fig.~\ref{transit_spin}, right). In three other systems in which the tidal interaction between star and planet is expected to be weak, no such discrepancy is observed \citep{2014A&A...565L...1P}. This points towards a possible spin-up, or inhibited spin-down, of stars which host tidally strongly interacting hot Jupiters. Other possible explanations, such as high-amplitude stellar activity cycles, are unlikely due to the observed constancy of the stellar X-ray emission level over several years \citep{2014ApJ...785..145P, 2011A&A...532A...3S, 2012AN....333...26P}. Observations of a larger sample of wide planet-hosting binaries are in progress.

\acknowledgments{
This work uses data from the X-ray Multi-Mirror Mission {\it XMM-Newton} and from the {\it Chandra} X-ray Observatory. K.P.'s work is performed under contract with the California Institute of Technology (Caltech) funded by NASA through the Sagan Fellowship Program.
}

\end{document}